# Study of a Prototype VP-PET Imaging System Based on highly Pixelated CdZnTe Detectors

Zheng-Qian Ye[1], Ying-Guo Li[1], Tian-Quan Wang[1], Ya-Ming Fan[1],
Yong-Zhi Yin[1,*], Xi-Meng Chen[1]

**Affiliations:**

[1]School of Nuclear Science and Technology, Lanzhou University, Gansu 730000, China

[*]Corresponding author. *E-mail address:* yinyzh@lzu.edu.cn

**Abstract:** We are investigating a prototype virtual pinhole positron emission tomography (VP-PET) system for small animal imaging applications. The PET detector modules were made up of 1.3 mm lutetium-yttrium oxyorthosilicate (LYSO) arrays, and the insert detectors consists of 0.6 mm pixelated Cadmium Zinc Telluride (CdZnTe). To validate the imaging experiment, we conducted a Monte Carlo simulation for the VP-PET system in Geant4 Application for Emission Tomography (GATE). For a point source of Na-22 with 0.5 mm diameter, the filtered back-projection (FBP) algorithm reconstructed PET image shows a resolution of 0.7 mm full-width-at-half-maximum (FWHM). The system sensitivity is 0.46 cps/kBq at the center of field view (CFOV) of the PET system with a source activity of 0.925 MBq and an energy resolution of 350-650 keV. A rod source phantom and a Derenzo phantom with F-18 were also simulated to investigate the PET imaging capability. GATE simulation indicated that two point sources with 0.5 mm diameter 2 mm apart could be clearly separated using 0.6 mm pixelated CdZnTe detectors as insert devices in a VP-PET system.

## 1 Introduction

Dedicated positron emission tomography (PET) scanners for small-animal studies have been investigated dramatically in the past decades. The imaging resolution of state-of-the-art small animal PET scanner is mostly limited to ~ 1 mm full-width-at-half-maximum (FWHM), which is not sufficient to carry out quantitative studies for the mouse brain. The targeted volume of the entire organ of mouse brain is as small as 0.5 cm$^3$. Thus, PET imaging system with sub-0.5 mm spatial resolution is required to achieve such kind of complicated mouse brain images. Extensive efforts have been focused on the improvement of the detector intrinsic spatial resolution using scintillation detector, semiconductor detector, gas detector, etc. Some novel PET geometries were also studied to improve the image resolution of the conventional PET scanners. One of such PET systems was called virtual-pinhole PET (VP-PET)[1-7], which has used high-resolution detectors integrated into a commercial PET scanner, in order to achieve both high resolution and high sensitivity.

CdZnTe detectors have been proposed as a high-resolution imaging detector candidate, with its root temperature operation, good spatial resolution, high energy resolution, and high detection efficiency for gamma rays[8-12]. With the difficulties of built application specific integrated circuits (ASIC) system for highly pixelated CdZnTe detectors, Monte Carlo (MC) simulation offers a cost-effective and useful method to understand the imaging capability of the PET based on pixelated CdZnTe detector[13]. Many studies were reported in aspects of both simulations and experiments for prototype conventional PET systems using CdZnTe detectors[14, 15]. Few reports were focused on the simulation of the PET insert system[16-17]. One of the challenges is the coincidence detection of two gamma-ray annihilation photons from an insert detector and PET scanner detector. The difficulty in mimicking the behavior of two photons generated from one prompt electron-positron annihilation in two kinds of detectors simultaneously is not solved. Researchers need to define the insert detector and the PET scanner detector simultaneously in the MC simulation, to track the two photons and calculate the energy deposition in both detectors[16-17].

GATE (Geant4 Application for Emission Tomography) is a widely used simulation software for emission tomography[18]. GATE has been used to validate many PET imaging systems for both animal studies and human clinics, including the system of ECAT EXACT HR+[19], ECAT HRRT[20], Hi-Rez[21], Allegro[22], GE Advance[23], MicroPET Focus 220[24], Inveon PET/SPECT/CT[25], Mosaic[26], Biograph mMR[27] etc. There were no published papers in which GATE software was used to validate PET insert system. GATE v7.0 has included the abilities to simulate triple-coincidence between two detection systems in one MC run. Researchers can define insert detectors integrated into the PET scanner, and analyze the photon hits in both PET scanner detectors and the insert detectors. In this paper, we defined LYSO detectors as PET scanner detectors and CdZnTe detectors as insert detectors in GATE, to build a VP-PET system.

## 2  Materials and Methods

### A. Small-animal PET configurations and Monte Carlo simulations

The proposed small-animal PET system was shown in Fig. 1. Two partial rings are included. The outer ring indicates PET scanner detectors, which are composed of 8 LYSO detector modules. Each LYSO detector is arranged in 18×18 crystal array. Every crystal element is molded into a square cross section of 1.2 mm × 1.2 mm with a 10 mm length. Gaps between elements are 0.1 mm width. Theoretically, the image resolution of this PET system would be better than the image resolution of Siemens Inveon PET scanner, which has LSO arrays with 1.59 mm crystal pitches[25].

The inner ring serves as insert detectors has 4 CdZnTe detector modules. Each CdZnTe detector consists of 16 ×16 pixelated elements. The pitch of CdZnTe detector is 0.6 mm, and the thickness is 5 mm. In GATE simulation, the software doesn't mimic the behavior of charge sharing of the CdZnTe detector. So, we select only the single-pixel photopeak events of pixelated CdZnTe detectors from the individual CdZnTe pixels.

The radius of the trajectories of the LYSO and CdZnTe detectors are 315 mm and 129 mm, respectively. The two detector modules are arranged into an asymmetric geometry. With the Virtual-Pinhole PET geometry, the projection of radioactivity distribution on the LYSO detector surface is magnified to 3.5 folds. Theoretical analysis indicates that the intrinsic spatial resolution of this prototype PET scanner is about 0.426 mm FWHM, and the system resolution is 0.956 mm FWHM[1].

In the real VP-PET experiments, we rotated the LYSO detectors and CdZnTe detectors to take cross-sectional data by step-and-shoot mode. An automated data acquisition system was developed using a rotation stage, a motion controller, and a PC with digitizers. CdZnTe detector and LSO detector was positioned at arbitrary distances from the center of field of view (CFOV) and rotated simultaneously to acquire coincidences from all possible angles to get a complete sinogram of a full-ring VP-PET insert system. Such setup would only allow us to record the coincidence events between CdZnTe insert detector and LYSO scanner detector (IS events).

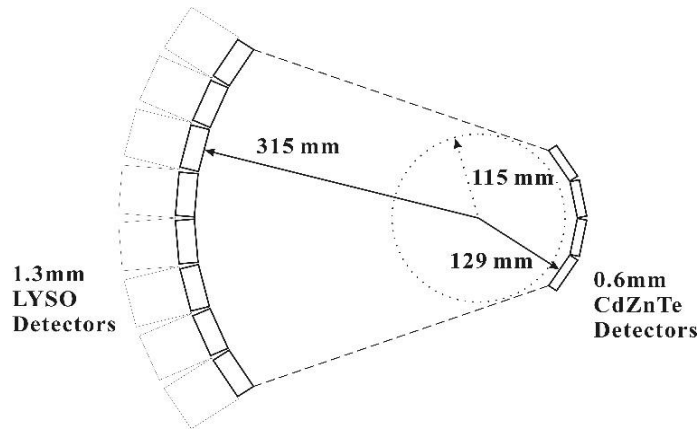

Fig.1 The experimental setup of the prototype PET Imaging system. Insert detectors are 600 μm pixelated CdZnTe detectors, and PET scanner detectors are LYSO arrays with 1.6 mm pitch. Not drawn to scale.

To simulate the performance of the prototype PET imaging system, we defined the PET system by repeated LYSO detectors and CdZnTe detectors into two whole rings, as shown in Fig.2. By this design, we are able to acquire three kinds of coincidence data. Those are coincidence events between LYSO detectors in PET scanner ring (SS events), coincidence events between CdZnTe detectors in insert ring (II events), and coincidence events between PET scanner detectors and insert detectors (IS events). The coincidence data of II and SS were recorded directly in GATE output files, while the IS coincidence data are calculated offline using the data sets of singles file output from GATE.

The physics process of the GATE simulation includes photoelectric absorption, Compton scatter, Rayleigh scatter, electron ionization, bremsstrahlung, and multiple scattering. An energy resolution of 20% and an energy window of 350-650 keV were defined, based on the LYSO detector measurements. We set a coincidence window of 20 ns and a time offset of 500 ns, due to the long electron drift time of CdZnTe detectors which we have measured in the real experiments[6]. We reconstructed the three coincidence data sets using a fan beam Filter-Back Projection (FBP) algorithm[28]. The positron source is F-18 with 10 kBq radioactivity. We also defined a rod source phantom and a Derenzo phantom in GATE simulation.

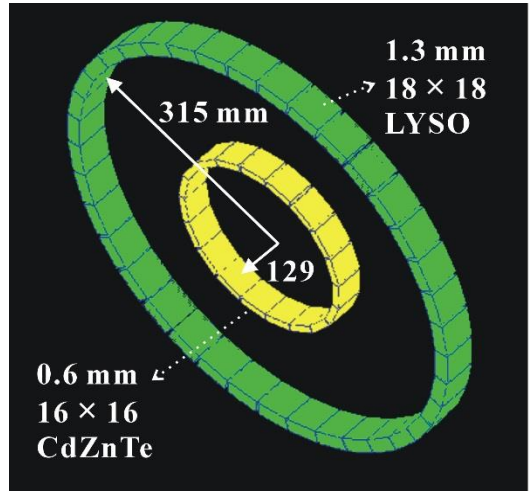

Fig.2 (Color online) GATE simulation geometry of the prototype PET Imaging system. The LYSO detectors and CZT detectors were repeated into two full rings for coincidence events acquisition.

B. **Spatial resolution and System sensitivity**

In order to obtain spatial resolution of this small-animal PET system, we imaged point sources throughout the region of interest (ROI). For the whole system size, effective detection range is set from 0 mm to 95 mm in trans-axial offset. The 0 mm indicates the center of Field of View (CFOV) of the PET system. Spatial resolution of the PET imaging system was measured in the radial offsets. A Na-22 point source in 0.25 mm diameter with 10 kBq radioactivity was stepped across the insert ring of the PET system. The radial offset of point sources selected in MC simulation was ranged from 0 mm to 42 mm, due to the limitations of FOV of the prototype PET scanner. From 0 mm to 30 mm radial offset, we move the point source by a 2 mm step. From 30 mm to 42 mm radial offset, we move the source by a 4 mm step, as shown in the horizontal coordinate of Fig.2. In total, 19 source locations were simulated and recorded by GATE. Coincidence events from the single-pixel photopeak events of pixelated CdZnTe detector were recorded, and the coincidence events from charge-sharing events of pixelated CdZnTe detector were rejected. PET image of point sources was reconstructed using a fan beam FBP algorithm[28].

System sensitivity is a measure of the annihilation photon detection capability of a PET system. For the VP-PET, the total system sensitivity would be higher than the system sensitivity of conventional PET scanners because of the integration of insert detectors. In this paper, Na-22 point source was used to calculate the system sensitivity of current prototype PET system. The radioactivity of the test point source was set to 0.925 MBq, and the energy window is 350-650 keV. For the pixelated CdZnTe detector, the collection of charge sharing events can improve system sensitivity of the PET system. We have demonstrated previously that if both double-pixel charge sharing events and single-pixel photopeak events were included into data acquisition process, the detection efficiency of the system would be increased 2.5 to 3 times[6]. But, with the limitations of GATE software, we could only record coincidence events from the single-pixel photopeak events of pixelated CdZnTe detector. In this real experiments, we will record both double-pixel charge sharing events and single-pixel events.

C. **Source phantom PET imaging**

Imaging performance of the PET system was characterized using point sources and phantoms filled with F-18. The point source of F-18 is a sphere with diameter of 0.5 mm, and the activity is 10 kBq. A Derenzo phantom with F-18 sources was also defined. The sources have diameters from 0.5 mm to 2.0 mm. We calculated the detecting matrix and reconstructed using FBP algorithm. The dimension of the detecting matrix hinges on the number of detector elements. On the X-Y plane the Insert ring has 1344 elements while the PET scanner ring has 1512 elements. During the GATE simulation, 200 thousand events were tracked. For source imaging data acquisition, we defined positron source as back-to-back gamma-ray mode. The choice of back-to-back gamma-ray sources was to speed up the MC simulation. By doing so, the positron range and acolinearity effects were not included in the MC study. If the positron range and acolinearity effect was excluded, the image resolution at CFOV would be 617 μm at FWHM.

3 **Results and Discussion**

A. **Spatial resolution**

Fig. 3 shows the radial spatial resolution of the PET system as the point source shifting along the radial direction. Both FWHM and FWTM resolutions were shown in the Fig.3. Radial resolution is getting worse when the source shifts to the edge of the PET system. It is verified that the radial resolution is affected by depth of interaction (DOI) of detection events from the PET system. The average FWHM of radial resolution obtained with the II, IS and SS coincidence events in the CFOV of the PET system were 0.74 mm, 0.83 mm and 1.24 mm, respectively. The FWTM of radial resolution obtained with the II, IS and SS coincidence data in the CFOV of the PET system were 1.38 mm, 1.58 mm, and 2.13 mm. In the whole range of insert ring, the radial resolution of the PET system reconstructed by the II, IS and SS coincidence data at FWHM was 0.63-0.87 mm, 0.78-0.99 mm, and

1.22-1.57 mm, respectively. While the radial resolution obtained by the II, IS and SS coincidence data at FWTM was 1.20-1.61 mm, 1.39-1.72 mm, and 1.99-2.19 mm, respectively.

As shown in Fig.3, the best image resolution in the radial directions were at the CFOV of the PET system. The spatial resolution would be decreased dramatically along the radial offset in the first 3 mm to 5 mm, and degraded much slowly when the source shifts to the edge of the PET scanner. The spatial resolution of II coincidence is the best one among all the three coincidences of II, IS and SS. The spatial resolution of IS coincidence is close to the spatial resolution of II coincidence, but is much worse than the spatial resolution of SS coincidence.

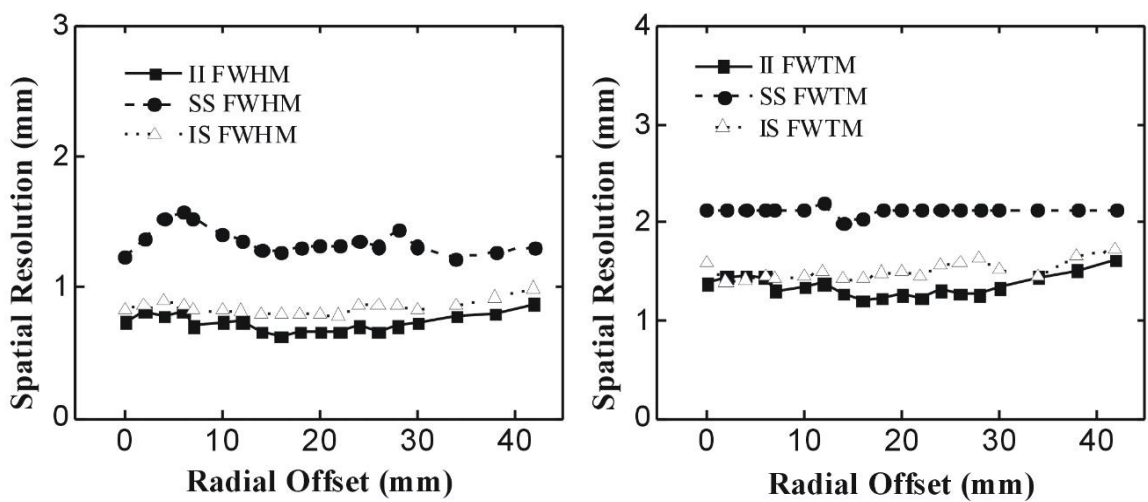

Fig.3 Radial resolution distribution through the PET system in the radial offsets, both FWHM and FWTM were shown.

B. **System sensitivity**

Fig. 4 shows the system sensitivity distribution in the axial directions, with an energy window of 350 keV - 650 keV. We only selected the gamma-ray detection within the insert ring range, i.e. from -7 mm to 7 mm. The whole width of the PET scanner ring is 23.4 mm along axial direction, while the whole width of insert ring is 9.6 mm. This makes the SS coincidence range wider than the II and IS coincidence range.

As shown in Fig.4, system sensitivity related to IS coincidence shows a triangle profile. In the axial range from -7 mm to -2 mm, the sensitivity of IS coincidence is lower than the sensitivity of SS coincidence, but is higher than the sensitivity of II coincidence. In the range from -2 mm to 2 mm, the sensitivity of IS becomes the biggest one among three kinds of coincidences. The maximum sensitivity of IS coincidence reaches 0.46 cps/kBq at the CFOV the PET system, given the point source activity of 0.925 MBq. This phenomenon could be well explained by higher gamma ray detection efficiency of CdZnTe detector, and shorter detection width of insert detector. The sensitivity of II coincidence is close to 0 at the edge of the PET scanner, and gradually increases in axial range from -7 mm to 0 mm. The maximum sensitivity of II coincidence is 0.18cps/kBq at the CFOV of the PET system. The sensitivity of SS coincidence fluctuates obviously between 0.23-0.35cps/kBq at the axial range of insert ring. The sensitivity of SS coincidence drops quickly when the source was placed towards the edge of the PET system.

In the CFOV of the system, the sensitivity of IS coincidence is the biggest one, and the sensitivity of II coincidence is the smallest one. This tendency keeps well agreements with the theory of VP-PET[1]. The spatial resolution and system sensitivity of our prototype PET indicated that the current design for small-animal PET imaging system based on IS coincidence is reasonable and feasible.

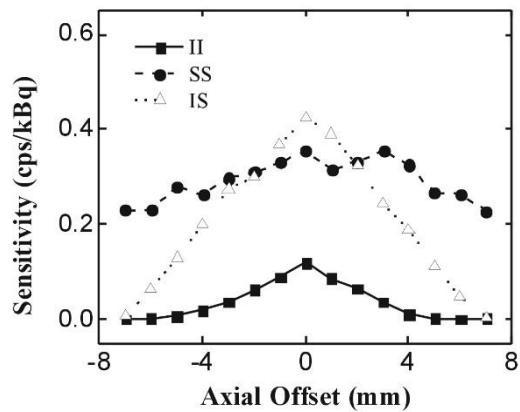

Fig.4 System sensitivity distribution in the axial directions, with an energy window of 350 keV - 650 keV.

C. **PET imaging of point source and phantom**

Fig. 5 shows the reconstructed image of an F-18 point source with a diameter of 0.5 mm placed at the CFOV of the PET system. Three reconstructed images for the II, IS and SS coincidence data were provided. The gamma-ray detection matrix for IS coincidence was also shown in Fig.5 (bottom right), which is a separate straight line for a point source. The axis of the detection matrix was corresponding to the crystal number of the insert ring and the PET scanner ring. Image resolution at FWHM for the II, IS and SS coincidence events were 0.49 mm, 0.7 mm and 0.88 mm, respectively. The image resolution at FWTM for the II, IS and SS coincidence events were 0.90 mm 1.26 mm and 1.62 mm. This result proved that the image resolution of IS coincidence of VP-PET is better than the image resolution of SS coincidence, which verifies the inserted high-resolution detector can improve the image resolution of the conventional PET system. The theoretical image resolution of IS coincidence is 1.195 mm at FWHM. The GATE simulation results keep good agreements with the theoretical calculations when positron range and acolinearity effect were included.

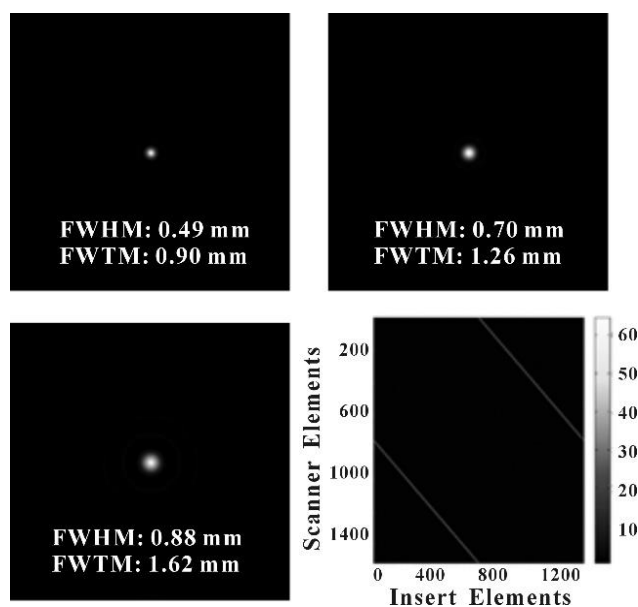

Fig.5 Reconstructed PET images of an F-18 point source with diameter of 0.5 mm for the II, IS and SS coincidence events (from top to bottom). The gamma-ray detection matrix of the IS coincidence events was also shown in the right bottom of the figure.

Fig.6 shows the reconstructed image using IS coincidence data for a rod source phantom and a Derenzo phantom. The diameter of the six point sources are 0.5 mm, 0.8 mm, 1.0 mm, 1.3 mm, 1.5 mm, and 2.0 mm respectively. The distance from each source to the center was 8 mm, and the distance between two sources was also 8 mm. It is noted that all the six point sources were clearly distinct. Fig.6 (right) shows the reconstructed image of a Derenzo phantom. The source dimension in the Derenzo Phantom remains the same with the source dimension in the rod source phantom. It is shown that the point sources of 0.5 mm in diameter 2 mm apart can be separated clearly.

GATE simulation indicates that the brightness of the point sources are decreased when the diameter of the sources increase. For the Derenzo phantom, when the radioactivity of the point sources are proportional to their diameter, the brightness of the reconstructed images of point sources would be displayed in the same gray scale. If the radioactivity of the point sources are changed as a square or cube of their diameter, the brightness of the reconstructed images of the point sources are successively weakened with their diameter increase. This phenomenon would be potentially explained by the partial-volume effect in PET system.

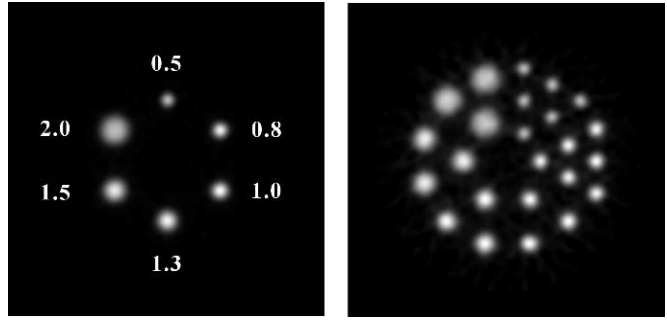

Fig.6. Reconstructed PET images of a rod source phantom and a Derenzo phantom for the IS coincidence events. The diameter of the sources were 0.5 mm, 0.8 mm, 1.0 mm, 1.3 mm, 1.5 mm, and 2.0 mm, from small to large.

D. **Discussion**

Virtual-pinhole PET (VP-PET) was initially defined and verified by the researchers in the Washington University in St. Louis (Tai, Wu et.al.), and sub-millimeter image resolution was obtained[1-3]. The studies of zoom-in PET were reported by the researchers in the University of California in Davis (Qi, Zhou et.al.)[4, 5]. In this paper, we calculated the image resolution of our prototype system using the equations of VP-PET[1], and very good agreements were found between MC simulated results and the theoretical calculations, as in Fig.7 (Left).

To compare the imaging resolution of the current PET system with the Micro Insert PET system designed in Washington University in St. Louis, we put the image resolution of MC simulated results and the real experimental results of Micro Insert system together, shown in Fig. 7 (Right). Micro Insert PET is the first VP-PET system[7], which integrated high-resolution LSO detectors into the MicroPET Focus-220. As we have reported in previous paper, the MC simulated results shows better image resolution than the real experimental results, due to (1) no electronic noise included, and (2) no positron range and acolinearity effect simulated for back-to-back gamma-ray sources. In Fig.7 (Right), the FWHM of SS, IS and II coincidence for the Micro Insert system was subtracted a factor of 0.5 mm, 0.4 mm and 0.2 mm respectively, because the acolinearity of PET system increased with diameter of PET scanner (SS > IS > II). After the correction, the image resolution of current PET system keeps very good agreements with the image resolution of Micro Insert system.

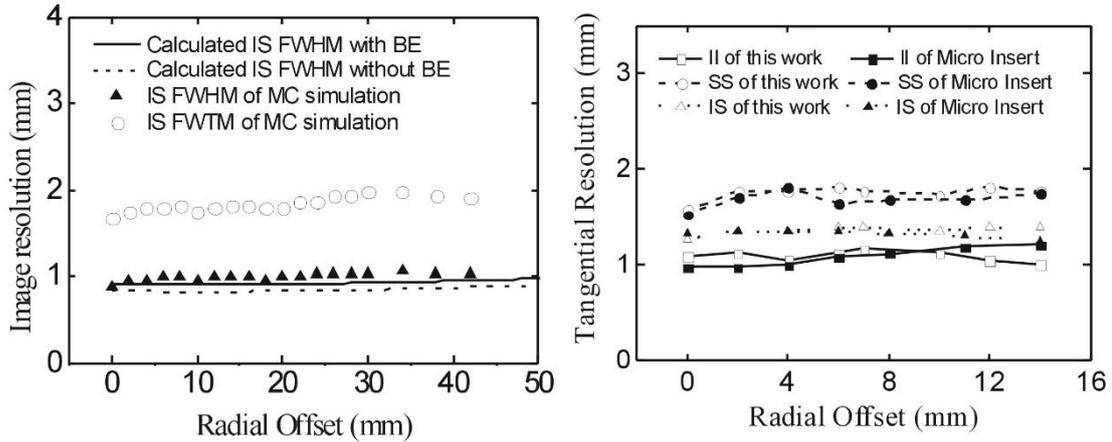

Fig.7 GATE simulated image resolution of the current prototype PET system was compared to the image resolution of the theoretical calculations (Left) and the image resolution of Micro Insert system [2] (Right).

## 4  Conclusion

In conclusion, a dedicated small-animal PET prototype system was simulated in GATE. The PET system has an inner and outer radius of 129 mm and 315 mm, which composed a virtual-pinhole structure to obtain high image resolution. The inner ring of the PET system consists of 600 μm pixelated CdZnTe detectors, meanwhile the outer ring of the PET system is composed of 1.3 mm LYSO detectors. Studies show that the system could be able to achieve PET images with a resolution of 0.7 mm at FWHM using FBP reconstruction algorithm. GATE simulation suggests that the radial resolution of the reconstructed image is within 0.74-1.00 mm at FWHM, and the tangential resolution ranges within 0.87-1.07 mm at FWHM. The system sensitivity at the CFOV of the PET system is 0.46cps/kBq (with source radioactivity of 0.925 MBq and energy window of 350 keV-650 keV). GATE simulation indicated that two point sources with 0.5 mm diameter 2 mm apart could be clearly separated using 0.6 mm pixelated CdZnTe detectors as insert devices in a VP-PET system.

**Acknowledgments**

Authors would like to thank the valuable discussions with Dr. Yuan-Chuan Tai at the Washington University in St. Louis, Dr. Heyu Wu in the Sinogram co. and Dr. Yujin Qi at the University of Wollongong. This work was supported by the National Natural Science Foundation of China (11305083) and the Fundamental Research Funds for the Central Universities under Grant lzujbky-2016-27.